\documentclass[conference]{IEEEtran}

\usepackage{cite}
\usepackage{amsmath,amssymb,amsfonts}
\usepackage{algorithmic}
\usepackage{graphicx}
\usepackage{textcomp}
\usepackage{xcolor}
\usepackage{xspace}
\usepackage{float}
\usepackage{amsmath}
\def\BibTeX{{\rm B\kern-.05em{\sc i\kern-.025em b}\kern-.08em
    T\kern-.1667em\lower.7ex\hbox{E}\kern-.125emX}}

\let\OldTextregistered\textregistered
\renewcommand{\textregistered}{\OldTextregistered\xspace}%

\let\OldTexttrademark\texttrademark
\renewcommand{\texttrademark}{\OldTexttrademark\xspace}%

\begin{document}

\title{Be Lean --- How to Fit a Model-Based System Architecture Development Process Based on ARP4754 Into an Agile Environment
}

\author{\IEEEauthorblockN{Daniel Dollinger, Julian Rhein, Kevin Schmiechen and Florian Holzapfel}
\IEEEauthorblockA{\textit{Institute of Flight System Dynamics} \\
\textit{Technical University of Munich}\\
M{\"u}nchen, Germany \\
\{daniel.dollinger, julian.rhein, kevin.schmiechen, florian.holzapfel\}@tum.de}
}

\maketitle

\begin{abstract}
An emerging service is moving the known aviation sector in terms of technology, paradigms, and key players - the Urban Air Mobility. The reason: new developments in non-aviation industries are driving technological progress in aviation. For instance electrical motors, modern sensor technologies and better energy storage expand the possibilities and enable novel vehicle concepts which require also novel system architectures for flight control systems. Their development is governed by aviation authority and industry recognized standards, guidelines and recommended practices. Comprehensive methods for Model-Based Systems Engineering exist which address these guidance materials but their setup and their application can be quite resource-demanding. Especially the new and rather small key players - start-ups and development teams in an educational environment - can be overwhelmed to setup such development processes. For these clients, the authors propose a custom workflow for the development of system architectures. It shall ensure development rigor, quality and consistency. The authors show how the custom workflow has been established based on the ARP4754A and its level of compliance to the standard's process objectives. Based on automation of life cycle activities, manual effort can be reduced to allow the application even in small teams. The custom workflow's activities are explained and demonstrated within a case study of an Experimental Autopilot system architecture.   
\end{abstract}


\section{Introduction}
This paper proposes a custom workflow derived from the ARP4754A~\cite{arp4754} as a Model-Based Systems Engineering (MBSE) approach for development of system architectures. It addresses clients such as small and agile development teams, usually found in an educational environment or start-ups. The custom workflow's objective is to ensure development rigor, quality, and consistency. This is achieved by focusing on essential process activities from the ARP4754A, selected objectives, their artifacts, and a high level of process automation.

The addressed clients are often involved in research and development activities targeting Urban Air Mobility (UAM) - an emerging service which moves the known aviation sector in terms of technology, paradigms, and key-players. Novel UAM vehicle concepts require novel system architectures for flight control systems to realize highly automated flight control functions and Simplified Vehicle Operation. Besides established design organizations, these start-ups and research institutions begin to participate in the race of this emerging aviation market. Their intention is to develop functional prototypes of such vehicles just for demonstrating their capabilities or even already with certification aspects in mind.

The development of aircraft and airborne systems is governed by aviation authorities. Industry recognized standards, guidelines, and recommended practices are used to demonstrate compliance with the certification regulations. However, their application can be challenging for these new key-players.

As existing methods for MBSE propose very comprehensive workflows to satisfy ARP4754A objectives, it can be resource-demanding to setup and conduct a consistent process which is in line with the recommended practice. Furthermore, not every development process activity is performed in the same tool which makes it even more complex to ensure traceability between artifacts (e.g. requirements, architectural models). As safety assessment is also a mandatory part of development, a well-defined and traceable interface to the safety assessment process and its artifacts with a clear input format is to be considered.


\subsection{State of The Art}
Instead of a traditional text based approach in documents, the development of system architectures can be supported by models. This so called Model-Based System Engineering (MBSE) is based on three pillars: the Domain Specific Language (DSL) which defines the model's framework, the tool to design models according to the DSL, and the method which describes the workflow of models and elements like process-activities \cite{Roques_ARCADIA}. In the best case, the tool already supports the method by giving advice on the workflow, proper automation, and model validation possibilities.

One of the very first modeling languages to help coping with complex structures was UML. It enables the specification, design and documentation of artifacts graphically\cite{UML}. These informal, box and line diagrams are generic diagram types which can be applied to many different engineering disciplines. However, these diagrams are standardized graphical notations with no discipline-specific knowledge. UML models can be validated according to UML constraints but extensions need to be setup first.

UML has been continuously developed until the current major release UML2 \cite{UML}. For UML2, different profiles are available which address specific engineering disciplines. A profile is an UML dialect to customize the language for its particular domain via stereotypes, tagged values, and constraints. One major outcome is SysML as a general purpose architecture modeling language. It supports specification, design, analysis, verification, and validation of systems from domains like hardware, software, data, etc. \cite{SysML}. However, SysML is based on UML which is mostly used in software development. This can be a disadvantage for system engineers who are not familiar with this discipline and haven't been exposed to object-oriented concepts \cite{Roques_ARCADIA, Mathew_MBSE}. MARTE is another UML2 profile which targets modeling of embedded real-time systems \cite{MARTE_UML_Profile} and therefore is a possible candidate to model particular elements of a system architecture. Still, MARTE primarily focus on embedded system's hardware / software modeling and the modeling-concept specification is quite comprehensive. 

When it comes to the second pillar - the tools for creating such architectural models - several solutions exist for the above mentioned languages. For instance, Microsoft Visio, MagicDraw, Rational Rhapsody, Papyrus, or MARTE Profile for Rational Software Architect. Some of them only provide a workbench for graphical editing whereas others already include features to validate models according to constraints and rules.

The ARCADIA framework specifies a DSL which is similar to SysML standards \cite{Batista_Capella}. It was introduced with a very broad scope of engineering disciplines. Additionally, ARACDIA provides also a method for MBSE of system, software, and hardware architectures. The method foresees different development steps. Some of them are: Operational Analysis, System Need Analysis, Logical Architecture, and Physical Architecture. A model is built for each step and justification links between each model ensure traceability and provide the possibility for impact analysis \cite{Voirin_Tales_MBSE}. The tool Capella is tailored to the ARCADIA method and supports the user in each of the method's development steps and the transition between them. It provides model checking rules for several categories and architects can define validation profiles for different aspects of the model \cite{Roques_ARCADIA}. With Capella, system architectures can be modeled with respect to artifacts stated in the ARP4754A. With custom interfaces to other tools or databases, traceability could be ensured between the requirements management and the safety assessment process \cite{Mathew_MBSE}. 
However, after the system architecture has been developed, hardware and software design proceed. In case of Model-Based Design (MBD), a tool has to be used which, in the best case, already supports development of software according to DO-178C \cite{RTCA178} and its supplements (eg. DO-331 \cite{DO331}). Having different tools for process activities, a big challenge is always to ensure traceability between development artifacts. To bridge the gap between system architecture development and software MBD, MathWorks introduced System Composer\texttrademark which is built on their MBD environment Simulink\textregistered \footnote{https://www.mathworks.com/products/system-composer.html}. With this approach, system architectures can be implemented in the same native environment as MBD would continue afterwards. The authors of \cite{Watkins_SystemComposer} show how they tailor their development environment in System Composer\texttrademark with profiles and stereotypes. They also proposed the possibility of behavioral modeling with an embedded Simulink\textregistered model inside System Composer.  

Based on this initial work, we want to continue and illustrate a custom workflow with System Composer\texttrademark which enables system architecture development from aircraft down to item level. Therefore, a DSL shall be developed which customizes System Composer\texttrademark to the aviation discipline. Relevant artifacts along the system development process model of ARP4754A shall be generated and a traceable connection to the safety assessment process activities as well as to requirements management and MBD shall be established.   

\subsection{Outline}
The paper is structured as follows. Section \ref{sec:ARP4754 Aircraft or System Development Process Model} describes the ARP4754A process structure and the selection concept of both, the process objectives and the process activities that shall be achieved and represented in the custom workflow. Within Section \ref{sec: Workflow Overview}, the custom workflow's activities are described. It shows the interrelation between each of them and illustrates the causal chain. Section \ref{sec: Automation of Life Cycle Activities} mentions the ideas of how to reduce effort in applying the custom workflow based on automation of life cycle activities, whereas section \ref{sec: Level of Compliance} shows the achieved compliance to the ARP4754A. A real world example as case study is presented in section \ref{sec: A Case Study}. It demonstrates the custom workflow's application for MBSE of an Experimental Autopilot. The paper concludes with section \ref{sec: Conclusions} and gives a short outlook into future research activities within \ref{sec: Future Work}.


\section{ARP4754A Aircraft or System Development Process Model}
\label{sec:ARP4754 Aircraft or System Development Process Model}

\subsection{Process Objectives and Selection Concept}
\begin{figure}	
	\centerline{\includegraphics[width=\linewidth]{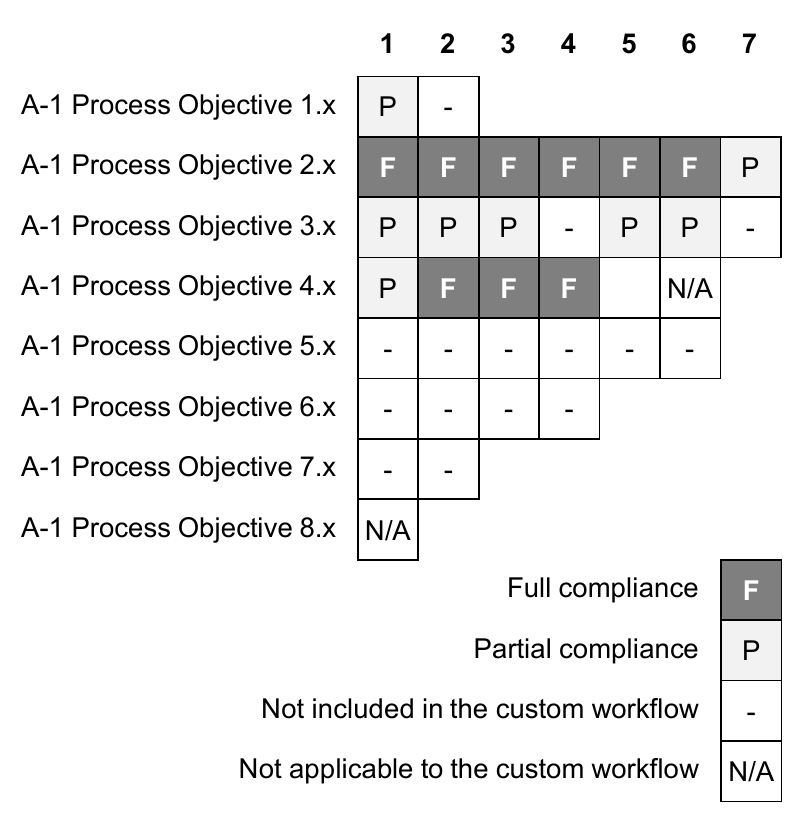}}
	\caption{ARP4754A Compliance Summary}
	\label{Fig:ObjectiveCompliance}
\end{figure}

Firstly, the objectives to be achieved when applying the custom workflow have to be selected. In the second step, the necessary process activities from the ARP4754A have to be identified which contribute to the desired objectives. A summary of the selected objectives from ARP4754A Table~\mbox{A-1} is shown in Fig.~\ref{Fig:ObjectiveCompliance}. It is based on following criteria:

\textbf{Development consistency and rigor -} is the primary focus. Objectives which output an artifact required in subsequent workflow activities have to be selected. The same holds for objectives which increase other artifact's development rigor with complementary information (e.g. safety requirements developed after conducting the Functional Hazard Assessment (FHA)).

\textbf{Traceability -} to achieve a traceable workflow from its entry level L0 to the lowest level possible of hardware and software requirements. Furthermore, traceability between different tools shall be guaranteed and gaps between workflow activities must not exist.

\textbf{Applicability to DAL level -} demonstrates the importance of an objective when it is recommended for certification even in lower DAL levels. However, this is only an indicator to prioritize objectives as no formal certification is to be achieved with the custom workflow.

\textbf{Interface to safety assessment -} is achieved by selecting objectives which output an artifact to be used as input to safety assessment activities and vice-versa. Therefore, the transition between the system development process and safety assessment process is clearly defined and internally developed safety assessment tools can be integrated (see section  \ref{section_safety_assessment}).

\textbf{Interface to implementation processes -} as final stage of the custom workflow. In order to conduct system implementation activities afterwards, objectives have to be selected which provide the necessary input to hardware and software development processes. For instance, the latter is addressed in \cite{Dimitriev_A_Lean_and_Highly} as a subsequent software development process to this custom workflow.

\subsection{Process Activities and Selection Concept}
\begin{figure*}	
	\includegraphics[width=\textwidth]{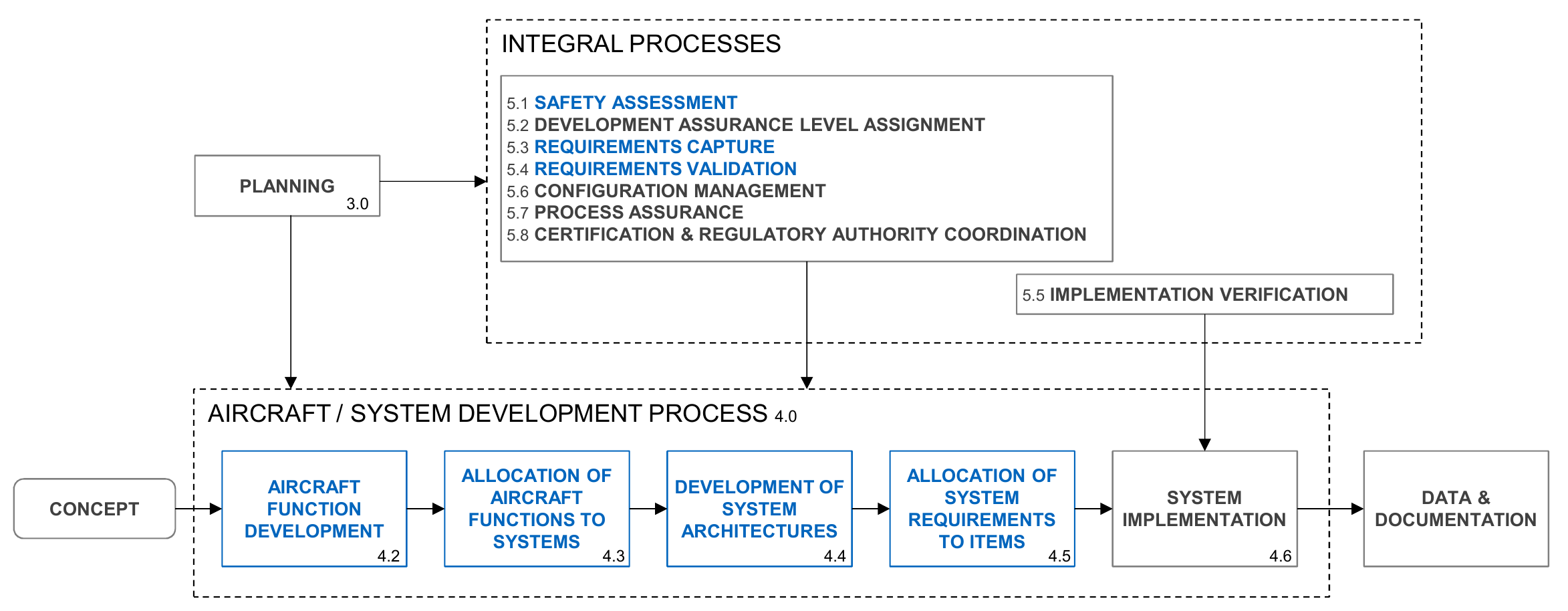}
	\caption{ARP4754A Aircraft or System Development Process Model (adapted from \cite{arp4754})}
	\label{Fig:ARP4754 Process Model}
\end{figure*}

The ARP4754A provides a generic aircraft and system development process model (see Fig.~\ref{Fig:ARP4754 Process Model}). It is further divided into three major processes: Development Planning, Integral Processes, and the Aircraft/System Development Process \cite{arp4754}. Each of them contains process activities related to process objectives and corresponding outputs, defined herein as artifacts.
Based on the previous selection of objectives, the related ARP4754A process activities have to be identified in order to be represented in the custom workflow and achievable by its MBSE approach. The below mentioned numbers correspond to the ARP4754A's chapter numbers where the process activities are described in. 

\textbf{Development Planning (3.0)} is not included in our workflow as its objectives primarily address creating documentation material serving as certification evidence. 

As some objectives from the \textbf{Integral Processes Safety Assessment (5.1)} have been selected, safety assessment activities are included in the custom workflow as well. A bidirectional interface to FHA and Preliminary System Safety Assessment (PSSA) is established. Artifacts from the development process are provided to the safety assessment activities and vice-versa such that a traceable connection between artifacts of both processes is established.

The \textbf{Integral Processes Requirements Capture/Validation (5.3, 5.4)} has been integrated to the custom workflow as requirements management activities. As an essential complement to the aircraft/system development process, consistent requirements authoring and documentation is mandatory on each system level. An interface to each architectural model is provided to accomplish the objectives of the requirements management activities. 

As the \textbf{Aircraft/System Development Process} is of overall interest to capture a system architecture, quite all of its process activities has been selected \textbf{(4.2 - 4.5)}. Based on this selection, consistency between development steps shall be achieved. Furthermore, development rigor shall be increased by taking feedback from safety assessment on a functional level. Therefore, relevant activities within the development process which provide the interface to the FHA must be taken into account. As a \textit{Physical Architecture} shall be the outcome of the custom workflow, such process activities need to be present as well. Finally, on the lowest hierarchical system level, requirements shall be allocated to software and hardware items which demands a corresponding process activity.

\begin{figure*}	
	\includegraphics[width=\textwidth]{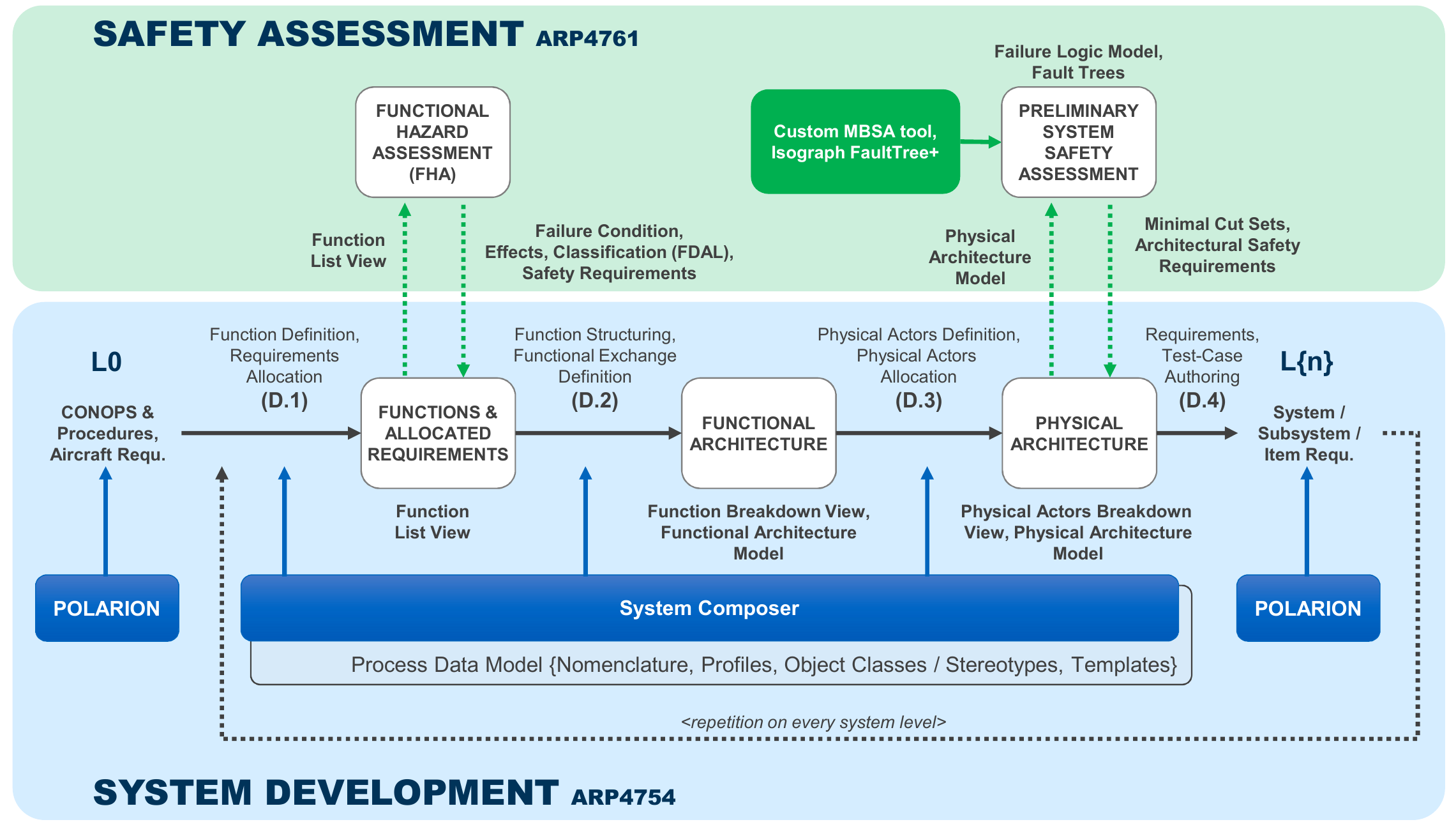}
	\caption{Custom Workflow in Relation to ARP4754A/ARP4761}
	\label{Fig:custom_workflow}
\end{figure*}

\section{Workflow Overview}
\label{sec: Workflow Overview}
\subsection{System Development Activities} 
To accomplish the system development activities, the tools Polarion\textregistered (see section \ref{section_requ_polarion}) and System Composer\texttrademark have been tailored to fit the custom workflow's need. For System Composer\texttrademark a process data model has been created which serves like a DSL to establish the modelling environment. For \textit{Functional} and \textit{Physical Architectures}, dedicated profiles contain stereotypes which enable the System Composer\texttrademark's components, ports, and connectors to represent an architectural item and its attributes for the corresponding architectural model. Examples are an \textit{Atomic Function} to be found in \textit{Functional Architectures} or an \textit{LRU} in \textit{Physical Architectures}. Modelling guidelines and templates further assist in creating the architectural models. Rules and constraints are in place to validate the model against.

Depending on the entry level from where the custom workflow is started, the input to this L0 (refer to Fig.~\ref{Fig:custom_workflow}) can either be the Concept of Operation (CONOPS), operating procedures, customer, and aircraft level requirements or system requirements from the next lower hierarchical level. Even an L0 at component level is possible.

With respect to the ARP4754A's process activity 4.2 (seen in Fig.~\ref{Fig:ARP4754 Process Model}), functions are developed by using System Composer\texttrademark components. A profile provides all necessary stereotypes to customize the modelling environment for \textit{Functional Architecture} development. Functional requirements are allocated to the functions from Polarion\textregistered with the internally developed and open source tool SimPol \cite{SimPol}. The output of this workflow activity D.1 is finally a function list which can be used in safety assessment activities (see section \ref{section_safety_assessment}). Feedback comprises for instance the Functional Design Assurance Level (FDAL) as a function's attribute as well as derived safety requirements authored in Polarion\textregistered and allocated to the respective function.

By structuring and/or logically grouping the developed functions, a \textit{Functional Architecture} can finally be established (D.2). Therefore, the \textit{Functional Exchange} between the functions has do be defined as System Composer\texttrademark's port and connector object. Based on the grouped functions, a \textit{Function Breakdown} view can be created using System Composer\texttrademark's built-in view functionality. It shows the architectural element's attributes and their hierarchical order.

The transition from functions to a physical system is done within the following workflow activity D.3. Design decisions are made, physical components are identified and selected. These physical actors which shall later fulfil the desired function, are placed within a \textit{Physical Architecture} model as System Composer\texttrademark components. Again, a profile provides all necessary stereotypes to customize the modelling environment for \textit{Physical Architecture} development. This includes for instance, stereotypes for LRUs, sensors or digital buses and their respective attributes. In large-scale models, the System Composer\texttrademark's built-in view functionality is very useful to create a \textit{Physical Actors Breakdown} view. Therein, subsystems and their physical components along with their attributes can be easily reviewed.

With respect to the ARP4754A's process activity 4.3, functions are allocated to the physical systems/components via System Composer\texttrademark's builtin model-to-model allocation functionality. As 4.3 is very specific on allocating aircraft functions to systems, the authors want to emphasize that the custom workflow can be applied to any hierarchical system level and therefore the expression of allocation is kept generic.

The process activity 4.4 in ARP4754A's process model is achieved by establishing interfaces between the physical actors in the \textit{Physical Architecture} model. Such interfaces are for instance digital communication buses, discrete signals, and power lines.

One iteration of the custom workflow concludes with authoring requirements in Polarion\textregistered for the physical actors present in the current \textit{Physical Architecture} hierarchical level (D.4). They are considered as input for the next workflow iteration cycle. On the lowest level L\{n\}, software and hardware item requirements are developed and can be used for instance as input to the development of electronic hardware according to DO-254 \cite{DO254} or model-based software development according to DO-178C/DO-331 \cite{Dimitriev_A_Lean_and_Highly, RTCA178}. The interrelation of and traceability between hierarchical levels in a \textit{Functional Architecture} as well as a \textit{Physical Architecture} model can be achieved by System Composer\texttrademark components with the stereotypes of \textit{Function Groups} or \textit{Subsystems} respectively. One could also use model references when desired.

\subsection{Requirements Management Activities} 
\label{section_requ_polarion}
\begin{figure*}	
	\includegraphics[width=\textwidth]{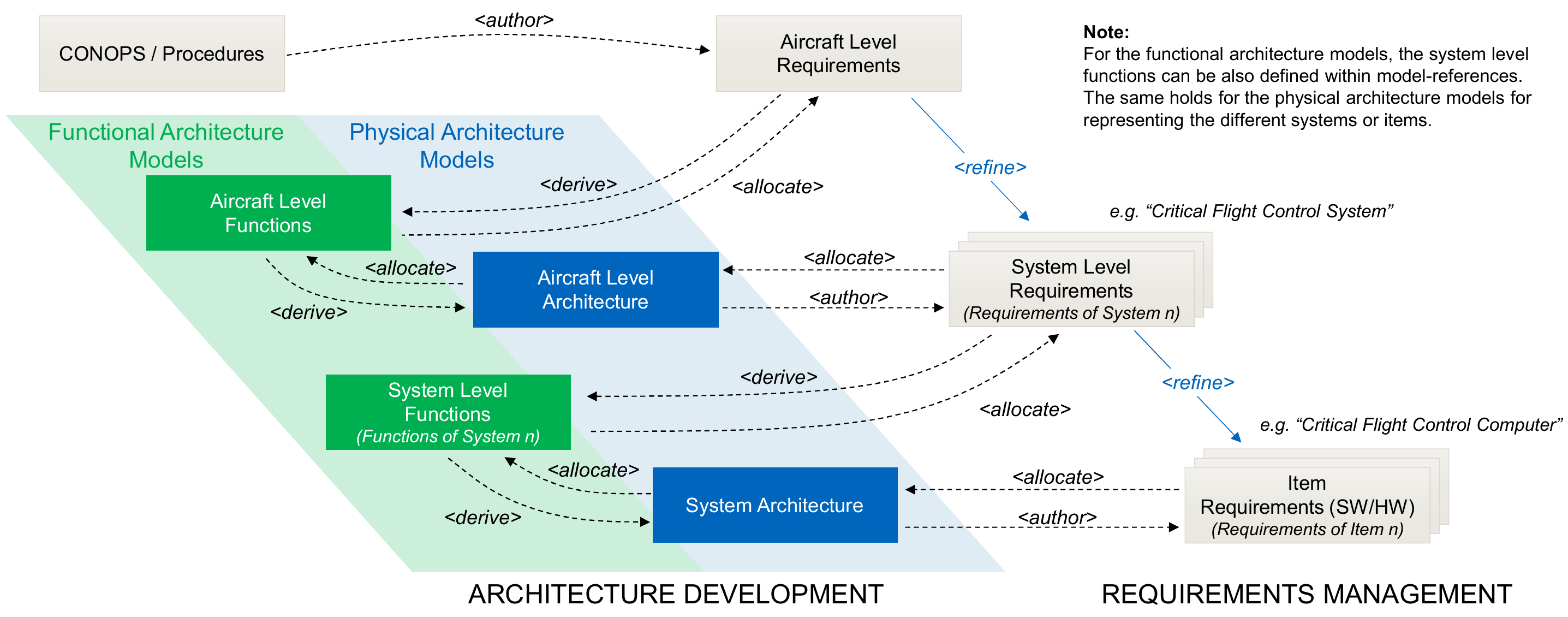}
	\caption{Custom Workflow - Architecture Development vs. Requirements Management Activities}
	\label{Fig:architecture_requirements_activities}
\end{figure*}

The interaction on each system level of the classical requirements management in Polarion\textregistered and architecture development in System Composer\texttrademark is depicted in Fig.~\ref{Fig:architecture_requirements_activities} for the complete process, starting at L0. It shows the sequence of development steps and not the linking relationship between the artifacts. 

From the textual CONOPS/Procedures and aircraft requirements created in Polarion\textregistered, the aircraft level functions are specified in System Composer\texttrademark in the custom workflow's step D.1. This satisfies the ARP4754A objective 2.1. The links between the two tools are created with SimPol. Step D.2 of the custom workflow creates the \textit{Functional Architecture}. In step D.3, the \textit{Physical Architecture} model at aircraft level is created and the aircraft functions are allocated to systems (objective 2.2). In the next phase, the system level requirements including assumptions are developed based on the requirements and the architecture of the aircraft level (objective 2.3, 2.4). These requirements are then further processed into system level functions and architecture in the System Composer\texttrademark (objective 2.5). 

In step D.4 of the custom workflow, the requirements and architecture on system level likewise lead to requirements on item (i.e. software and hardware) level (objective 2.6). The form of those software requirements depends on the approach that has been chosen for the software development process according to \cite{RTCA178, DO331}. For the classical approach, they are textual in Polarion\textregistered, however, for the model-based approach (example 4 and 5 of \cite{DO331}), they are also model-based and as referenced Simulink\textregistered model part of the System Composer\texttrademark environment.

With this process, it is possible to establish complete traceability across tools in the whole chain from CONOPS down to software and hardware item requirements in order to satisfy objective 4.4. With the traceability between the requirement levels, we consider objective 4.1 partially fulfilled because missing requirements or conflicting requirements are noticed with a high probability while establishing the links between the artifacts.

Assumptions are not shown in Fig.~\ref{Fig:architecture_requirements_activities} but are also managed on Polarion\textregistered. Objective 4.2 is satisfied by the justification available with the assumptions and the linked test cases used to validate them. The derived requirements, which are not based on higher-level requirements but on architecture decisions are identified as such by their link to the \textit{Physical Architecture} artifacts. The justification for the derived requirement is stored with the artifact and linked test cases are used to validate it. This fulfills objective 4.3.

\subsection{Safety Assessment Activities}
\label{section_safety_assessment}
As safety assessment is an integral part of the systems development process according to \cite{arp4754}, it is only natural to built safety assessment support into the custom workflow presented in this article. The goal of this integration is to enable reusability of the artifacts from the system development process (i.e. functions, items, physical connections, etc.) for safety assessment purposes and to ensure traceability and consistency between both processes.

Technically, this is implemented by a custom toolset, which enables the import of the functional and physical architecture into the respective safety modeling environment, where those structural models are enriched with safety related information. In the case of the functional architecture, the set of functions is imported and extended by failure conditions and effects for each function during the FHA. 

For the analysis of the physical architecture, a custom model-based safety assessment (MBSA) solution is employed, which allows to reuse the physical architecture as a so-called fault propagation model (FPM). Within the FPM, the structure of the physical architecture model is extended by sets of internal failures, propagated failures and their logical connection based on the principle of component fault trees (see \cite{BernhardKaiser.2003}). Furthermore, a bi-directional linking mechanism enables navigation from the system architecture models to the associated FPM. Consistency between the system architecture models and the safety assessment models is ensured by a synchronization functionality, which updates the FPM accordingly when the linked system architecture model changes. For a more detailed discussion of the technical principles refer to \cite{Rhein.2021}. 

The results of the safety assessment process are then available again for the system development, however formal traceability or import functionality is not part of the current process. While those approaches in principle support the objectives of the safety assessment process according to \cite{arp4754}, they do not fully automate them or enforce completeness of the required analyses. Consequently, the level of fulfillment is classified as partial.

\section{Automation of Life Cycle Activities} 
\label{sec: Automation of Life Cycle Activities}
Automation of life cycle activities shall reduce the manual effort in applying the custom workflow. Therefore, the focus has been set to identify essential key drivers as for instance time-consuming manual reviews. Based on validation rules, such reviews can be automatically executed as so called model validation tasks. They have been divided into process and modelling guidelines compliance rules and their corresponding validation. Furthermore, development effort can be reduced not just by detecting issues during reviews on the final implementation stage but also during the process through immediate and repetitive feedback of the validation results. The framework is designed such that it allows also the definition of new rules by the system architect.

\subsection{Process Validation Rules}
Process validation rules shall ensure that the development has been conducted according to the underlying workflow. It can be referred to as process assurance. For instance, these rules are designed to detect breaks in traceability, deviation from nomenclatures or templates. The latter two comprise an automated review on consistent use of stereotypes within the architectural models whereas the former  evaluates requirements allocation or missing stereotype parameters. Additionally, missing connections of architectural elements could demand justification by the system architect. 

\subsection{Model Validation Rules} 
Model validation rules shall ensure the correct application of modelling guidelines and the constraints they enforce. This is mainly achieved by evaluation based on the assigned stereotypes.

\begin{figure}
	\centerline{\includegraphics[width=0.75\linewidth]{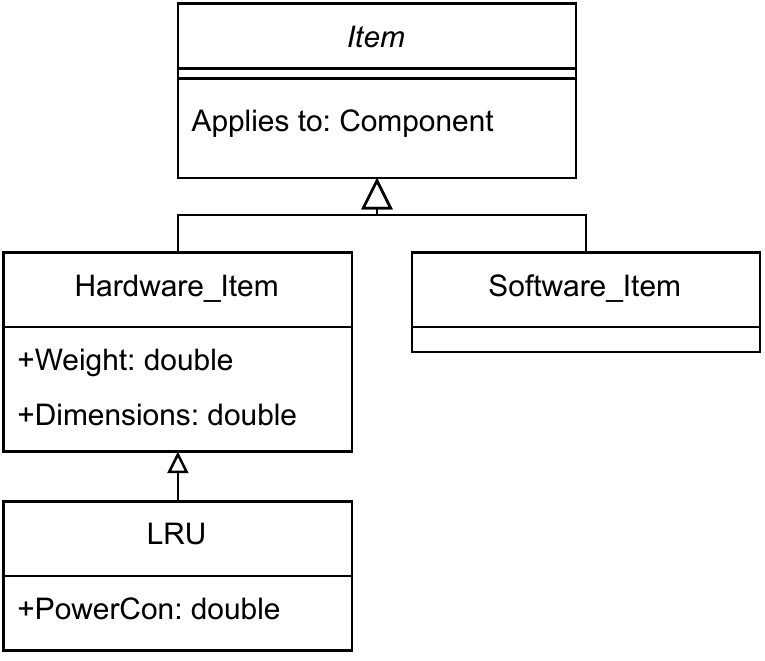}}
	\caption{Stereotypes Related to Components and Items (excerpt)}
	\label{Fig:class_diagram_components}
\end{figure}

\begin{figure}
	\centerline{\includegraphics[width=0.75\linewidth]{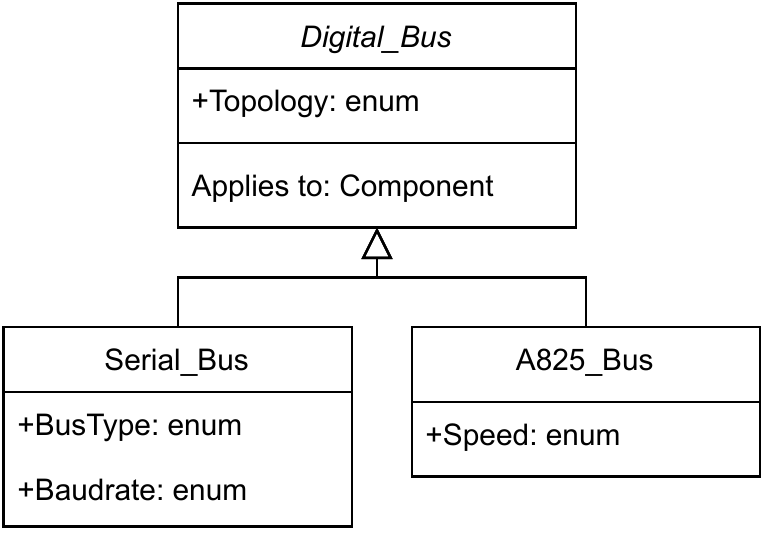}}
	\caption{Stereotypes Related to Digital Busses (excerpt)}
	\label{Fig:class_diagram_digital_bus}
\end{figure}

\begin{figure}
	\centerline{\includegraphics[width=0.75\linewidth]{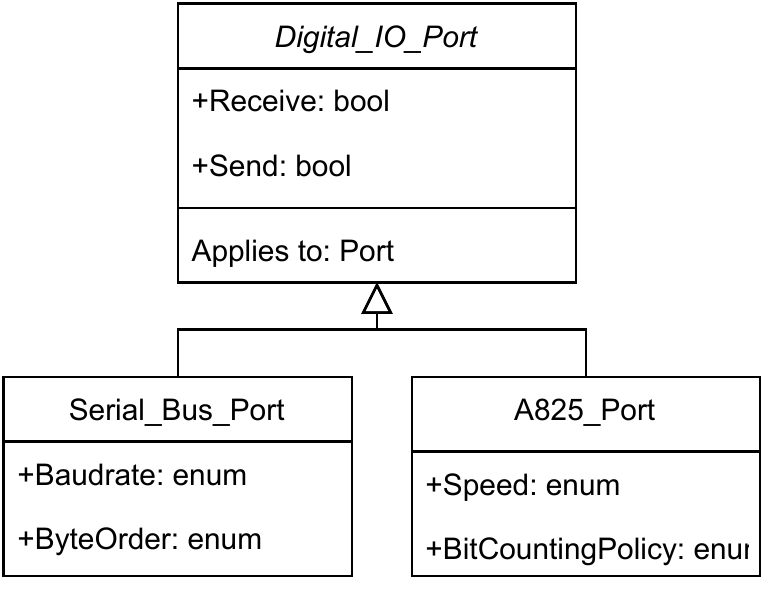}}
	\caption{Stereotypes Related to Digital Ports (excerpt)}
	\label{Fig:class_diagram_ports}
\end{figure}

For each type of architectural model, a dedicated profile exists with some predefined stereotypes. They have been initially setup based on the aviation sector's terminology within MBSE. Class diagrams which contain a subset of the \textit{Physical Architecture} profile and its stereotypes is depicted in Fig. \ref{Fig:class_diagram_components} to Fig. \ref{Fig:class_diagram_ports}.

New stereotypes can be created by the system architect also by inheriting existing base-stereotypes. Therewith, parameters of the base-stereotype are inherited as well. As mentioned before, new rules can be created to evaluate modelling constraints of new stereotypes. For instance, such constraints shall ensure that an ARINC825 port within the \textit{Physical Architecture} is only connected to another ARINC825 port.

\section{Level of Compliance}
\label{sec: Level of Compliance}
The ARP4754A specifies various process activities to be executed - some even with independence - in order to achieve its objectives and thereby generating recommended evidence to be used in the certification liaison process. Fig.~\ref{Fig:ObjectiveCompliance} contains a summary of the ARP4754A process objectives and the authors assumption to which level of compliance they are satisfied by the proposed custom workflow:
\begin{itemize}
    \item Full compliance - all activities are represented in the custom workflow to produce the objective's artifacts
    \item Partial compliance - not all activities are represented and not all artifacts are produced
\end{itemize}
Note that when it comes to independence, the authors assumed that this will be handled individually by the applicant.

Following activities are omitted as no formal certification by authorities is intended when applying the custom workflow:
\begin{itemize}
	\item The ARP4754A's planning process and its activities to establish development plans. The authors assume that in the addressed environment, overall project planning is considered to be sufficient.
	\item At the authors institute, a lightweight configuration management (CM) was designed in GitLab for development arteficts and in Polarion\textregistered for documents and requirement baselines \cite{Dimitriev_A_Lean_and_Highly}. However, the custom workflow does not contain any particular CM activities.
	\item Dedicated process-assurance activities are not included as well as no activities for the certification and regulatory authority coordination process.
\end{itemize}

\section{A Case Study} 
\label{sec: A Case Study}
As for a pilot project, the custom workflow has been applied on system architecture development of an Experimental Autopilot with L0 being at system level. The Experimental Autopilot itself has been developed in previous studies such as \cite{DA42Autopilot} and \cite{DA42Development} at the Institute of Flight System Dynamics. Based on the existing example system, the MBSE approach shall demonstrate that the custom workflow, together with its automated life cycle activities, proofs development rigor, quality and consistency.  

As initial input, a subset of system level requirements are authored in Polarion\textregistered. The \textit{Functional Architecture} model is being developed based on a System Composer\texttrademark template, provided in the custom workflow's framework. It is part of a MATLAB\textregistered project and already contains a profile with the basic stereotypes for a \textit{Functional Architecture}. This profile is available through the model standards as part of our developed DSL. 
According to the custom workflow's activity D.1, the system level functions are defined. These functions describe what the system shall do in order to satisfy the requirements - firstly in a solution agnostic manner. 
In this paper, we focus on the subset of functions which are derived from the certification requirement that it must be possible to disconnect the experimental autopilot and allow manual control of the aircraft. Therefore, the function \textit{Engage/Disengage Autopilot Surface Control} is defined as a component block in System Composer\texttrademark and the stereotype \textit{Atomic Function} is selected. With the stereotype, certain attribute fields are added to the component block and provide the possibility to add for instance a function description, rationale, and FDAL. With SimPol, the relevant requirements are linked using the Simulink\textregistered Requirement Management Interface (RMI) \cite{SimPol}. 

\begin{figure}	
	\centerline{\includegraphics[width=0.7\linewidth]{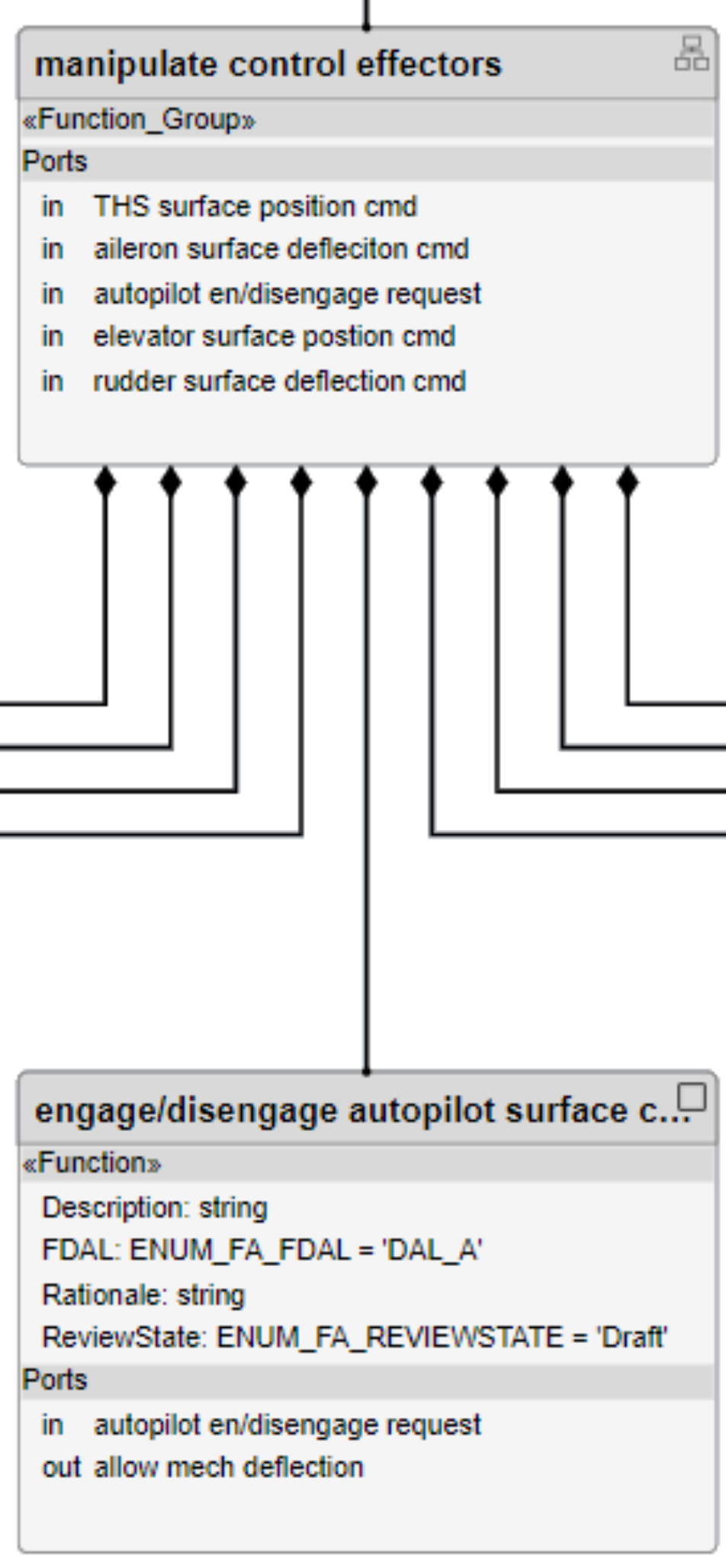}}
	\caption{System Composer\texttrademark - Architecture View as Function List}
	\label{Fig:functional_breakdown}
\end{figure}

In the template model, System Composer\texttrademark Architecture Views are already configured to display the model's content as a \textit{Functional Breakdown} (see Fig.~\ref{Fig:functional_breakdown}). Besides \textit{Atomic Functions}, the stereotype \textit{Function Group} can be allocated to a component block such that it serves as a logical grouping of functions. As the \textit{Functional Breakdown} is a graphical representation, it can only be used for manual and visual review purposes. In order to provide a \textit{Function List} as an artifact to the FHA tool, the \textit{Functional Architecture} model and its architectural elements are translated into custom MATLAB\textregistered classes. Therewith, all architectural content is available in a more generic model and interfaces can be established to safety assessment or other tools.

As output from the FHA, safety requirements are authored in Polarion\textregistered and the FDAL attribute is provided to the functions in System Composer\texttrademark. It has been identified that \textit{Engage/Disengage Autopilot Surface Control} can lead to a severe failure condition with the effect being that manual control is not possible if the Experimental Autopilot cannot be disengaged. A catastrophic failure effect classification leads to a resulting FDAL assignment.  

\begin{figure*}
	\includegraphics[width=\textwidth]{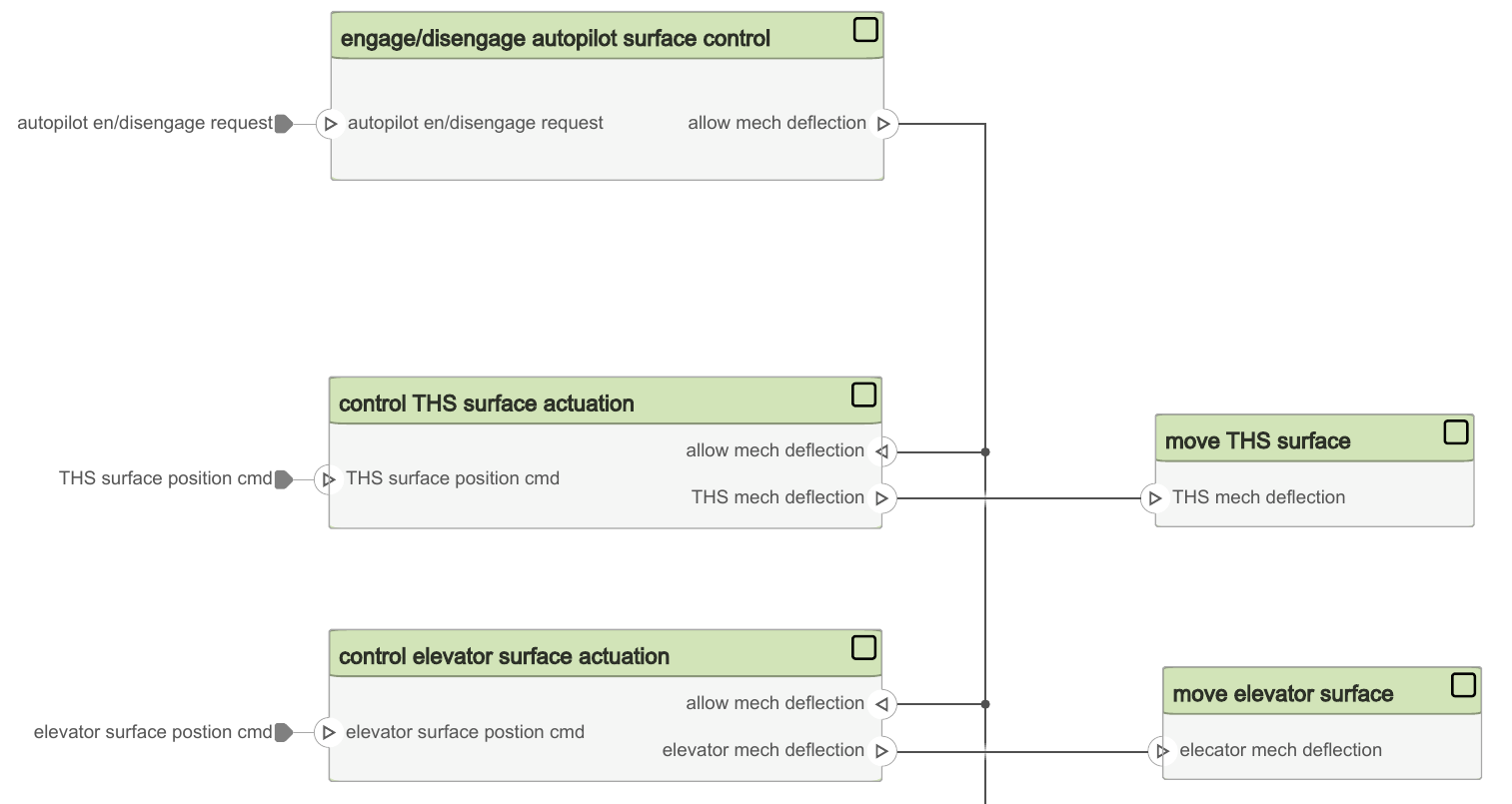}
	\caption{System Composer\texttrademark Functional Architecture Model (excerpt)}
	\label{Fig:functional_architecture}
\end{figure*}

\begin{figure*}	
	\includegraphics[width=\textwidth]{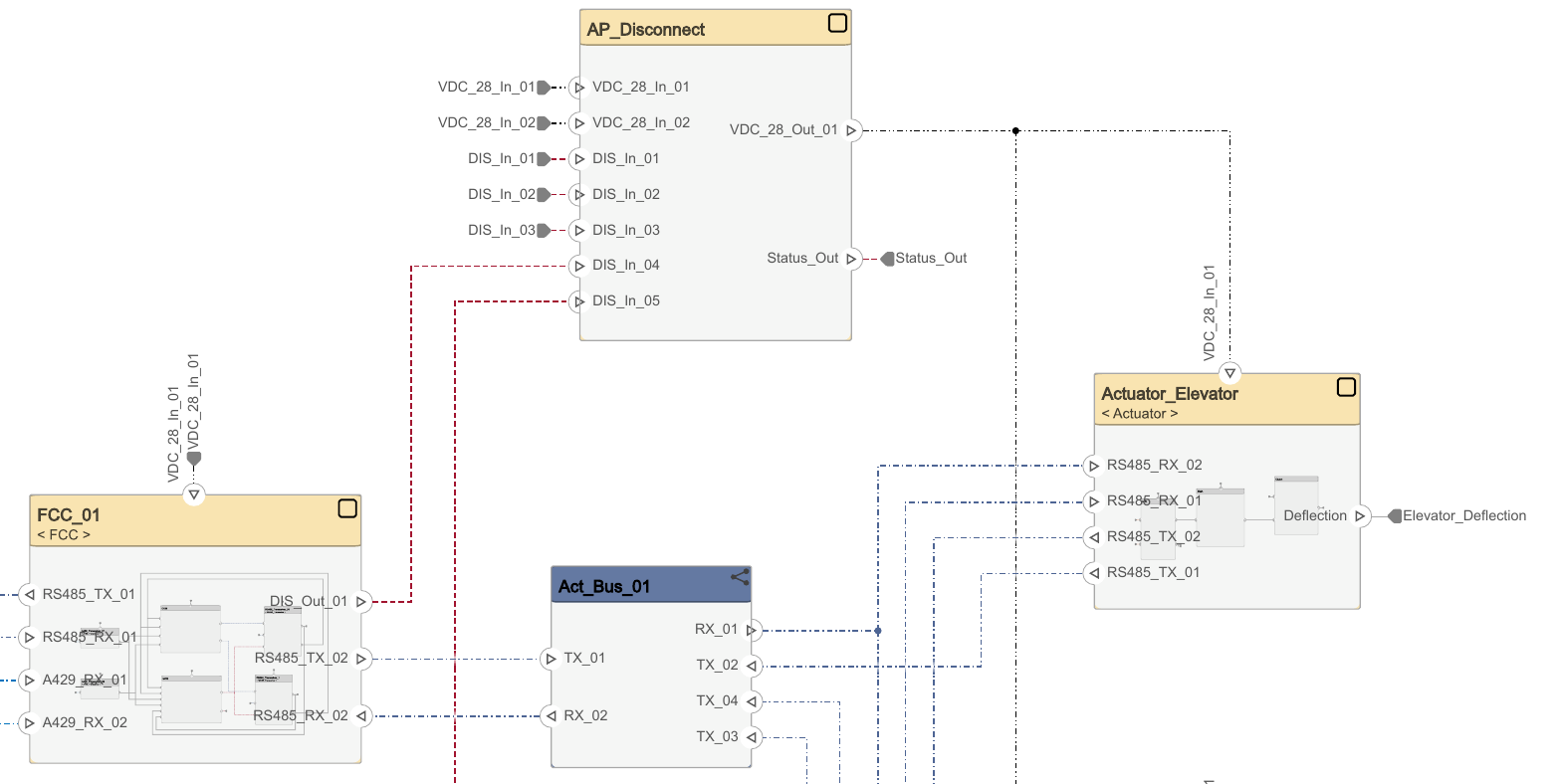}
	\caption{System Composer\texttrademark Physical Architecture Model (excerpt)}
	\label{Fig:physical_architecture}
\end{figure*}

The previously mentioned generic model based on MATLAB\textregistered classes is also used to perform checks against the process and model validation rules. As an example, the class \textit{Atomic Function} is checked whether a requirement has been allocated to all of its instances or not. 

In the custom workflow's activity D.2, the \textit{Functional Exchange} is added between \textit{Atomic Functions}. For instance, the function \textit{Engage/Disengage Autopilot Surface Control} receives an \textit{autopilot engage/disengage request} from another function and outputs \textit{allow mechanical deflection} to the flight control surfaces as seen in Fig.~\ref{Fig:functional_breakdown}. In completing all exchanges, the artifact \textit{Functional Architecture} is created (see Fig.~\ref{Fig:functional_architecture}). 

D.3 starts with design decisions of how to realize the previously defined functions. This includes the identification of required physical components - the so called physical actors definition. A System Composer\texttrademark template is also available for the \textit{Physical Architecture} model and already contains a profile with stereotypes for common physical components (see Fig.~\ref{Fig:class_diagram_components}). In an analogous way as Fig.~\ref{Fig:functional_breakdown} is showing, a \textit{Physical Actors Breakdown} can be created with System Composer\texttrademark's \textit{Architecture Views}. 

The function \textit{Engage/Disengage Autopilot Surface Control} shall be implemented by the physical component \textit{AP\_Disconnect}. To show this relationship and thereby ensure traceability, the System Composer\texttrademark's \textit{Allocation Editor} is used to link the function within the \textit{Functional Architecture} model to the physical component within the \textit{Physical Architecture} model.

Where necessary, a connection between physical components has to be establish in order to realize the \textit{Functional Exchange}. In comparison to the OSI model, this connection represents the physical layer. For this purpose, available stereotypes can be allocated to System Composer\texttrademark's ports and connectors. Such can be a digital data bus, a discrete signal or a power supply. In Fig.~\ref{Fig:physical_architecture}, a discrete signal is received by the \textit{AP\_Disconnect} component which represents the \textit{autopilot engage/disengage request}.

Again, based on the generic model which is derived from the System Composer\texttrademark content, checks can be performed against the process and model validation rules. For instance, it is being checked whether the discrete port \textit{DIS\_In\_04} of \textit{AP\_Disconnect} seen in Fig.~\ref{Fig:physical_architecture} is connected only to another discrete port. In this case \textit{DIS\_Out\_01} from \textit{FCC\_01} .

In order to interact with the Preliminary System Safety Assessment (PSSA), the \textit{Physical Architecture} model and its architectural elements are translated into custom MATLAB\textregistered classes as well. Based on failure logic models and fault trees, minimal cut sets can be determined with which the system architecture and the design decisions behind can be validated against the safety requirements.

One iteration of the custom workflow concludes with activity D.4. Requirements are developed and authored in Polarion\textregistered for instance to the HW and SW items of the physical component \textit{AP\_Disconnect}. With SimPol, a link can be established back to System Composer\texttrademark in order to ensure traceability. 

\section{Conclusions}
\label{sec: Conclusions}
The goal of this paper was to propose a custom workflow as an approach for MBSE in the context of ARP4754A which shall ensure rigor, quality and consistency during the development without being as resource-demanding as existing methods. By selecting only a subset of ARP4754A process objectives as well as process activities, a first contribution to this goal was made. Besides that, the custom workflow ensures consistency throughout the development and does provide complete traceability of artifacts from L0 to L\{n\}.

With regard to the three pillars of MBSE, the custom workflow itself provides a sophisticated method whereas modelling guidelines, modelling constraints, and templates act like a DSL. The tool System Composer\texttrademark has been customized with profiles and stereotypes to act as the third pillar, together with Polarion\textregistered, as an implementation environment. With automation of life cycle activities, another step in reducing resource-demanding efforts has been made. Furthermore, these activities also increase development rigor as well as quality through design checks against process and model validation rules. 

As L0 for the custom workflow can be set at various hierarchical system levels, the output of the lowest level L\{n\} are HW and SW item requirements where the latter can serve as input to the subsequent process of MBD according to DO-178C/DO-331 \cite{Dimitriev_A_Lean_and_Highly}. As section \ref{section_requ_polarion} describes, SW requirements can also be model-based and for instance be a referenced Simulink\textregistered model within a System Composer\texttrademark component with the stereotype \textit{Software\_Item} (see Fig.~\ref{Fig:class_diagram_components}). Therewith, the custom workflow's abilities extend the currently present concepts to link between system architecture development and MBD.  

\section{Future Work}
\label{sec: Future Work}
With future studies, the authors want to extend the interaction between MBSE and MBD. Where MBD is mostly used for application SW development with automatic source code generation, low-level SW source code for hardware interfaces and the interaction with the application SW is usually manually written. By including the (OSI) transport layer into a \textit{Physical Architecture} model, together with specification models (refer to DO-331), overall system simulations could be conducted within System Composer\texttrademark. In a following step, also the low-level SW source code for the embedded target's hardware interfaces could be automatically generated based on this transport layer definition.


\section*{Acknowledgment}
The authors want to express their gratitude to Prof. Dr.-Ing. Stephan Myschik for his time during all discussions and his great support as reviewer.

\bibliographystyle{ieeetr}
\bibliography{references}

\end{document}